\documentclass{article}

\usepackage{arxiv}
\usepackage{amsmath}

\usepackage[utf8]{inputenc} % allow utf-8 input
\usepackage[T1]{fontenc}    % use 8-bit T1 fonts
\usepackage{hyperref}       % hyperlinks
\usepackage{url}            % simple URL typesetting
\usepackage{booktabs}       % professional-quality tables
\usepackage{amsfonts}       % blackboard math symbols
\usepackage{nicefrac}       % compact symbols for 1/2, etc.
\usepackage{microtype}      % microtypography
\usepackage{lipsum}		% Can be removed after putting your text content
\usepackage{graphicx}
\usepackage{natbib}
\usepackage{doi}
\graphicspath{{images/}}

\hypersetup{
colorlinks=true,
linkcolor=black
}
\hypersetup{
citecolor=black
}

\title{A WT-ResNet based fault diagnosis model for the urban rail train transmission system}

\author{ 
    Zuyu Cheng\\
	School of Mechanical Engineering\\
    Xi’an Jiaotong University\\
    Xi’an 710049, China\\
	\And
    Zhengcai Zhao\\
	School of Mechanical Engineering\\
    Xi’an Jiaotong University\\
    Xi’an 710049, China\\
    	\And
    Yixiao Wang \\
	School of Mechanical Engineering\\
    Xi’an Jiaotong University\\
    Xi’an 710049, China\\
    \And
    Wentao Guo \\
	School of Mechanical Engineering\\
    Xi’an Jiaotong University\\
    Xi’an 710049, China\\
    \And
    Yufei Wang \\
	School of Mechanical Engineering\\
    Xi’an Jiaotong University\\
    Xi’an 710049, China\\
    \And
    Xiang Gao \\
	School of Mechanical Engineering\\
    Xi’an Jiaotong University\\
    Xi’an 710049, China\\
}

\renewcommand{\shorttitle}{A WT-ResNet based fault diagnosis model for the urban rail train transmission system}

\hypersetup{
pdftitle={A WT-ResNet based fault diagnosis model for the urban rail train transmission system },
pdfsubject={q-bio.NC, q-bio.QM},
pdfauthor={David S.~Hippocampus, Elias D.~Striatum},
pdfkeywords={First keyword, Second keyword, More},
}

\begin{document}

\maketitle

\begin{abstract}
	This study presents a novel fault diagnosis model for urban rail transit systems based on Wavelet Transform Residual Neural Network (WT-ResNet). The model integrates the advantages of wavelet transform for feature extraction and ResNet for pattern recognition, offering enhanced diagnostic accuracy and robustness. Experimental results demonstrate the effectiveness of the proposed model in identifying faults in urban rail trains, paving the way for improved maintenance strategies and reduced downtime.
\end{abstract}

% keywords can be removed

%第一部分 引言
\section{Introduction}
In March 2021 at the meeting of the Commission for Financial and Economic Affairs under the CPC Central Committee, Carbon peaking and neutrality were seriously taken account of and incorporated into the overall plan of China for environment protection, which has proposed higher demand of the quantity and quality of urban railway system. The progression of urban railway system are reflected in the safety and reliability of transportation vehicles. Therefore, ensuring the normal operation of these vehicles has emerged as a crucial issue confronting the current development. Under the circumstances, the conventional methods and modes of the operations and maintenance of vehicles are no longer adequate to meet the current demand that they have to operate with high efficiency, rigorous standardization and precise refinement.[\cite{he2020research}]

The conventional method of operations and maintenance entails conducting maintenance activities only upon the detection of a fault in the mechanical equipment, which is called “fireman” as they just can come in handy after the accidents happened, putting the operations and maintenance in a passive state [\cite{XDXK201812008}].Gradually, a method of regular diagnosis emerged, which involves assessing the operational status of mechanical equipment periodically to determine whether the maintenance is required or not. However, it faces issues related to timeliness, as guaranteeing the timely detection and resolution of faults is definitely a tough work. It’s common to miss the accident when adopting the regular diagnosis pattern.

The advancement of modern information technology and artificial intelligence (AI), exemplified by technologies like big data analysis, machine learning, and deep learning algorithms, has led to the emergence of intelligent operations and maintenance as a highly effective approach for equipment upkeep. By employing methodologies such as data mining, association analysis, and trend analysis, the development of an intelligent operation and maintenance system for equipment is facilitated. This system offers sophisticated features such as remote monitoring, fault detection, expert diagnosis, health management, and other intelligent functionalities, which serve to improve the efficiency, accuracy, reliability, and versatility of operation and maintenance processes, while simultaneously reducing labor costs and minimizing operational errors [\cite{BFJD201903012}].

Intelligent fault diagnosis is an essential part of intelligent operation and maintenance, which has made great, advanced progress and achievements with the development of AI. The traditional machine learning methods, such as support vector machine(SVM) and artificial neural network(ANN), have been widely used in fault diagnosis, but they have limited ability to process natural data and rely to much on the expertise in some specific area. Presently, deep learning(DL) has been widely used and made tremendous contributions to the fault diagnosis technology. DL theory was first published in Science in 2006 by Hinton et al[\cite{hinton2006reducing}] .It performs outstandingly in discovering intricate structures in high-dimensional data[\cite{lecun2015deep}] ,  which is beneficial to some classification or detection tasks, hence, DL is utilized widely in industry and science. DL methods, such as stacked self-encoders(SAE), convolutional neural network(CNN) and recurrent neural network(RNN), are studied and used widely in fault diagnosis. SAE is a type of artificial neural network used for unsupervised learning. It consists of multiple layers of autoencoders, where each layer learns to encode the input data and reconstruct it at the output. Trained layer by layer using some techniques like backpropagation, SAE can perform well in the tasks like feature learning and dimensionality reduction. CNN is a supervised learning network, aiming to learn spatial hierarchies of features from input data automatically. It consists of diverse layers, including convolutional layers, pooling layers, and fully connected layers, and are very useful in facial recognition and autonomous vehicles. RNN is designed for process sequential data by retaining the previous inputs. A fundamental aspect of RNN is their capability to discern dependencies and patterns within sequential data through the context of preceding inputs. Based on its features, RNN is commonly used in machine translation, natural language and time series analysis. 

DL has progressed greatly these years, the residual neural network(ResNet) is one of the achievements. ResNet was first proposed by Kaiming He et al. to address the degradation problem[\cite{he2016deep}] , which made it easier to train deeper neural networks and optimize them. It’s better for ResNet to converge more easily when trained, helping reduce the time cost for training process. Moreover, ResNet performed well in generalization ability, allowing it to adapt to multiple sets of data and tasks while performing outstandingly in terms of accuracy and efficiency. In this article, we utilize the ResNet to construct the system of diagnosis and analyze its performance.

%第二部分 
\section{Theoretical foundation}
\subsection{Convolutional neural networks}
\begin{figure}[h]
	\centering
	\includegraphics[width=\textwidth]{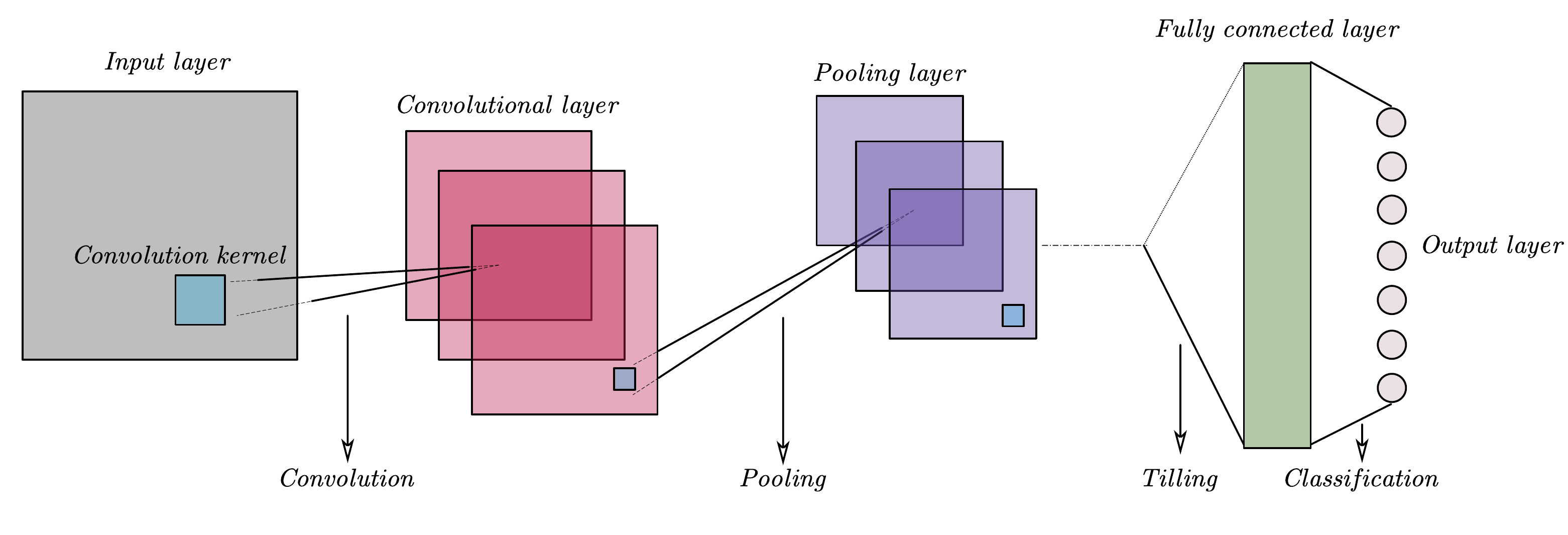}
	\caption{Convolutional neural network structure diagram}
	\label{fig:CNN}
\end{figure}

Convolutional Neural Network(CNN) was first proposed by LeCun and initially applied to image recognition. Due to the characteristics of local receptive fields, shared weights, and spatial subsampling, CNN has been successfully applied in various fields such as document recognition, speech recognition, spectrum recognition, and fault diagnosis[\cite{726791}].The structure of CNN typically consists of multiple alternating convolutional layers, pooling layers, and fully connected layers. The convolutional and pooling layers are responsible for extracting features layer by layer, while the fully connected layers are responsible for feature classification or regression. A typical CNN structure is  shown in Figure \ref{fig:CNN}.

\paragraph{Convolutional Layer: } The convolutional layer plays a crucial role in the feature extraction process by applying convolution operations to the input or the features from the preceding layer to generate fresh features that serve as the input for the subsequent layer. This operation is underpinned by the convolutional kernel filter, which is instrumental in capturing the local features present within the input data. By sliding across the input data, the filter produces a series of outputs, known as feature maps, where each position corresponds to a different feature activation. The convolutional kernel filter functions as a shared weight mechanism, linking the current layer to the next layer, a concept referred to as weight sharing. Subsequent to the convolution operation, a non-linear activation function is applied to the resulting feature map to augment its expressiveness. This entire convolutional process is mathematically encapsulated in formula \ref{euq:1}, which formalizes the transformation of the input data into a more abstract feature representation.

\begin{equation}
X_k^l=\phi\Bigg(\sum_{k=1}^KW_k^l*X_k^{l-1}+B_k^l\Bigg)
\label{euq:1}
\end{equation}

Here,$X_k^l$ represents the $k-th$ output of convolutional layer $l$, where $k$ denotes the number of convolutional kernels in layer $l$, $W_k^l$ represents the $k-th$ convolutional kernel of layer $l$, $*$ denotes convolution operation, $X_k^{l-1}$ represents the $k-th$ output of the layer $l-1$, $B_k^l$ represents the bias of convolutional layer $l$, $\phi$ represents the non-linear activation function applied to the convolution output. Non-linear activation functions can map features to non-linear spaces, thereby enhancing the linear separability of features. Common non-linear activation functions include the Sigmoid function, Tanh function, ReLU function, etc.

\paragraph{Pooling Layer:} Positioned following the convolutional layer, the pooling layer serves a dual purpose: it not only distills the essential local information from the extracted features but also reduces the dimensionality of these features, thereby decreasing the computational demand associated with the subsequent parameter updates. Among the available pooling techniques, max pooling stands out as the predominant choice due to its ability to select the maximum value within a specified region, as mathematically expressed in equation \ref{euq:2}, which effectively captures the most prominent feature within that area.

\begin{equation}
P_{ijk}^{l}=\max\left(x_{ijk}^{l}:i\leq i<i+p,j\leq j<j+q\right)
\label{euq:2}
\end{equation}

Here, $P_{\ddot{v}k}^{l}$ represents the output of pooling layer $l$, $x_i\dot{v}^{l}$ represents the$ (i, j)$ element in the $k-th$ feature map output by convolutional layer $l$, and $p$ and $q$ respectively represent the length and width of the pooling window.

\paragraph{Fully connected layer:} Once the input data has undergone the initial processing steps of convolutional and pooling layers to extract relevant features, it enters the fully connected layer where these features are transformed into a one-dimensional form suitable for classification. This layer typically consists of several hidden layers, culminating in an output layer that performs the actual classification. The output layer often employs the Softmax function to convert raw scores into probabilities, with each class probability computed using the formula shown in equation \ref{euq:3}, thus completing the classification process.

\begin{equation}
O_j=
O_{j}=\left[\begin{array}{c}
P(y=1 / x ; \theta \\
P(y=2 / x ; \theta \\
\cdots \\
P(y=k / x ; \theta
\end{array}\right]=\frac{1}{\sum_{j=1}^{k} \exp \left(\theta^{j} x\right)}\left[\begin{array}{c}
\exp \left(\theta^{1} x\right) \\
\exp \left(\theta^{2} x\right) \\
\cdots \\
\exp \left(\theta^{k} x\right)
\end{array}\right]
\label{euq:3}
\end{equation}

Where $k$ represents the number of categories, and $\exp(\theta^{j}x)$ represents the parameters of the classification layer.

\subsection{The residual network}
\begin{figure}[h]
	\centering
	\includegraphics[width=0.7\textwidth]{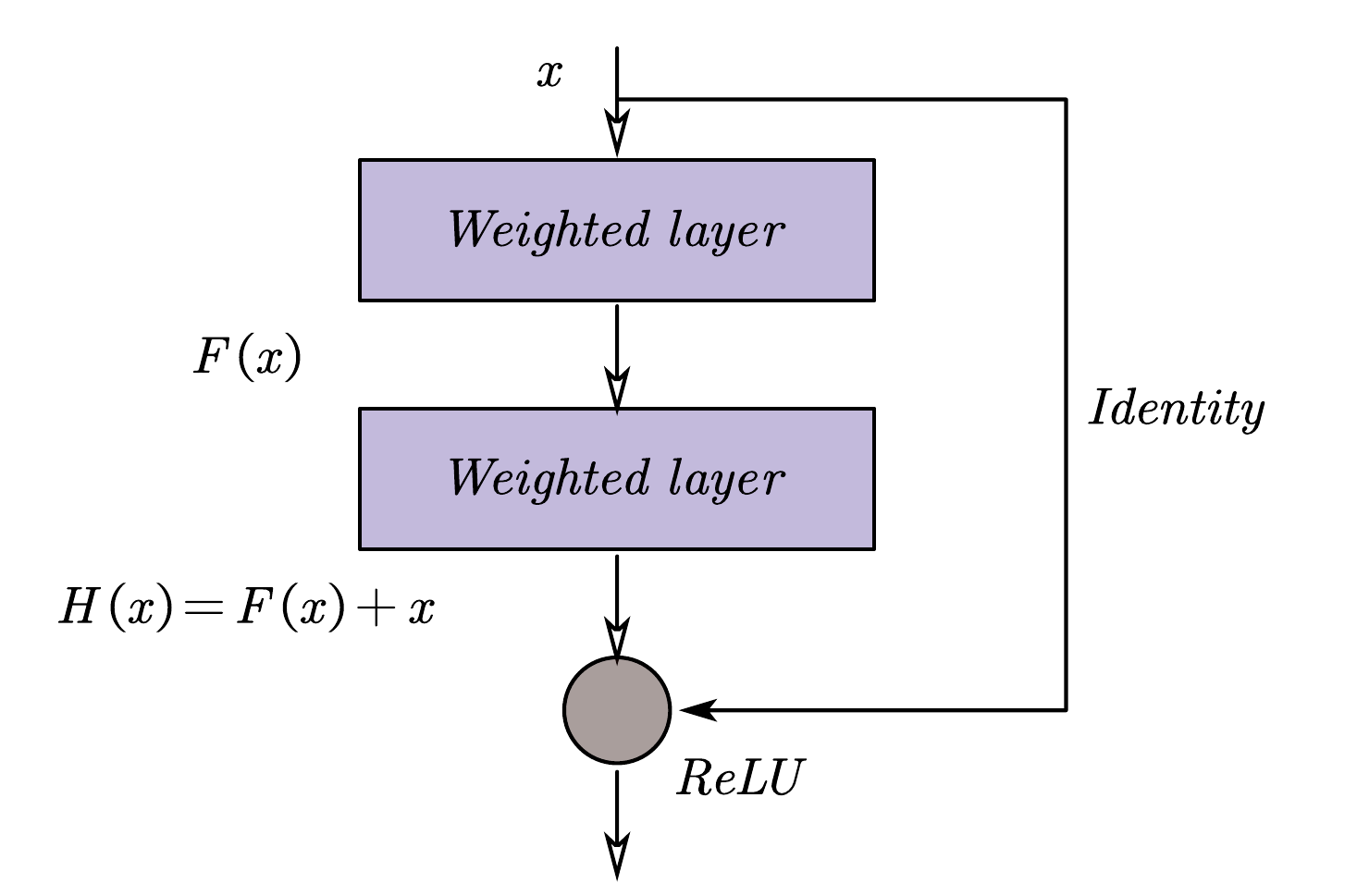}
	\caption{Structure diagram of the residual block}
	\label{fig:Resnet_block}
\end{figure}

In 2016, He et al. made a breakthrough in deep learning by introducing the Residual Neural Network (ResNet)[\cite{2016Deep}], which was designed to overcome challenges such as model degradation and overfitting that often arise in deeper neural network models. ResNet's innovative approach involves the use of residual blocks that facilitate the learning of residual features, thereby enhancing the flow of information through the network. This groundbreaking architecture achieved remarkable success in the ImageNet image recognition competition, setting new standards in the field of computer vision. Meanwhile, other notable neural network models, such as AlexNet[\cite{2012ImageNet}], GoogLeNet\textbf{}, and VGGNet[\cite{2014ImageNet}], had already made significant impacts in the early to mid-2010s, but they were generally considered shallow network models. These models, while successful in their own right, faced limitations when attempting to scale to greater depths. Residual neural networks introduced the concept of residual learning, learning residual features through multiple connected residual blocks, as shown in Figure \ref{fig:Resnet_block}.

Here, $x$ is the input of the residual neural network, $H(x)$ is the output, $F(x)$ is the residual mapping function.  $H(x)=F(x,\{W_i\})+x$ is the identity mapping function, and the $Weight layer$ is the convolutional layer. The residual neural network adds cross-layer fitting of residual functions on the basis of ordinary deep convolutional networks, only learning the difference between output and input. He et al. demonstrated through experiments that fitting the residual mapping function $F(x)=H(x)-x$ is much easier than fitting an identity mapping function $H(x)=x$[\cite{2016Deep}].During the training process, lower-layer errors can be passed to the upper layer through shortcut connections, adding residual gradients in addition to using the target function gradient. Therefore, while the residual neural network has a deeper layer, it also has stronger feature learning capabilities.

\begin{figure}[h]
	\centering
	\includegraphics[width=\textwidth]{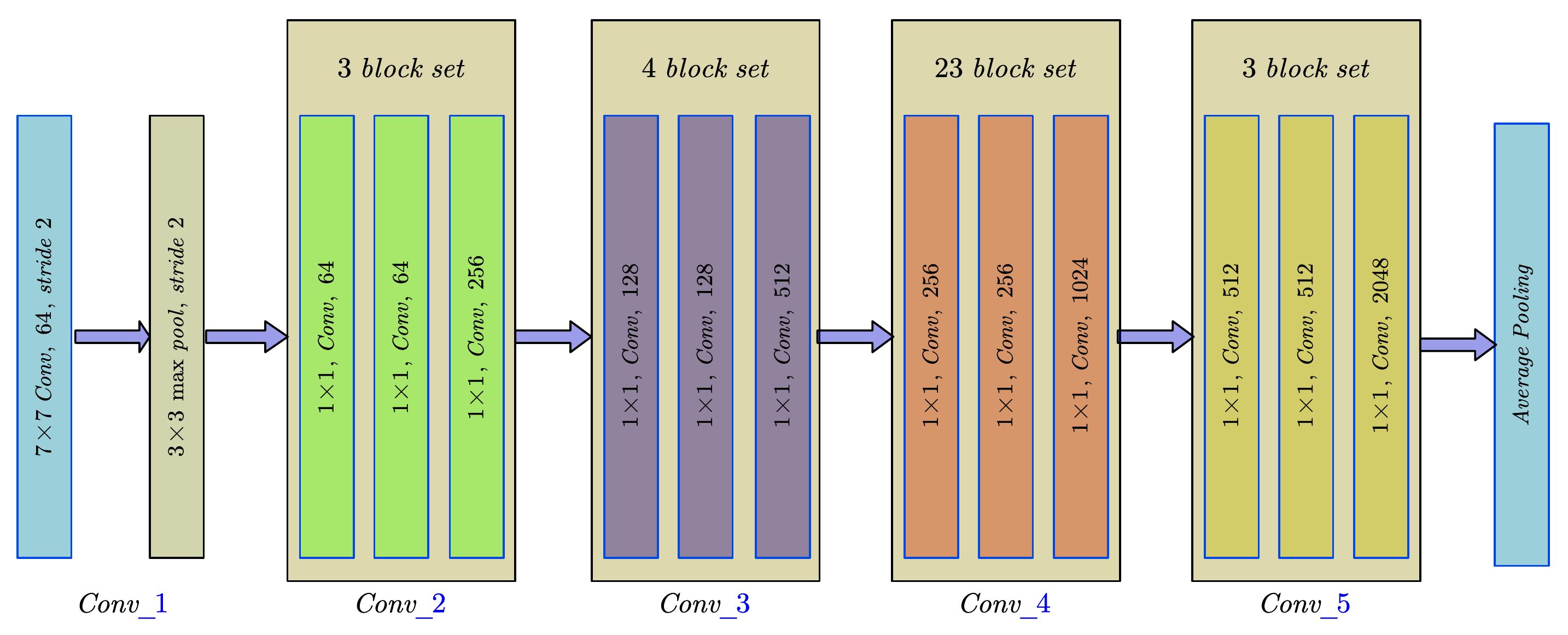}
	\caption{ Resnet-101 network diagram}
	\label{fig:Resnet}
\end{figure}

The residual neural network structure typically consists of two pooling layers, multiple residual blocks, and fully connected layers. The depth of the network can be adjusted by stacking different numbers of residual blocks. Common residual network models of different depths include Resnet-18, Resnet-34, Resnet-50, Resnet-101, etc[\cite{2016Deep}]. Among them, the Resnet-18 network structure is shown in Figure \ref{fig:Resnet}

In Figure \ref{fig:Resnet}, when the number of channels in the current and previous layers is the same, the residual connection is an identity mapping, marked with solid lines in the figure. When the number of channels in the current and previous layers is different, a $1*1$ convolution is needed to change the number of channels in the residual connection, maintaining the same dimensions for both layers.

When the number of channels in the current and previous layers is the same, the output of the residual block is:
\begin{equation}
y=F(x,W)+x
\label{euq:4}
\end{equation}
When the number of channels in the front and back layers is different, the residual connection needs to set a $1*1$ convolution $W_s$ to change the number of channels, keeping the dimensions of the front and back consistent. This is indicated by a dashed line in the diagram. At this point, the output of the residual block is:
\begin{equation}
y=F(x,W)+W_sx
\label{euq:5}
\end{equation}
By using the outputs of residual blocks from formulas\ref{euq:4} and \ref{euq:5}, we can obtain the output of the $l-th$ layer of a residual neural network.
\begin{equation}
y_l=x_l+F(x_l,W_l)
\label{euq:6}
\end{equation}
Pooling is another important operation, usually used to reduce the dimension of features. There are many methods for performing pooling operations, such as max pooling or average pooling, as shown in formulas \ref{euq:7}and\ref{euq:8} .

\begin{equation}
z_{ij}^{k}=\max_{pq}^{k}\left(p,q\in R_{ij}\right)
\label{euq:7}
\end{equation}

\begin{equation}
z_{ij}^{k}=\frac{1}{|R_{ij}|}\sum_{pq}^{k}(p,q\in R_{ij})
\label{euq:8}
\end{equation}

Here, $z^k_ij$ represents the output in the $k-th$ feature map of the pooling operation; $x^k_pq$ represents the value of the neuron in the pooling region $R_{ij}$.

%%\lipsum[4] See Section \ref{sec:headings}.

\section{Data cleaning}
The datasets provided by the commitee are collected from the experimental platform, where multi-source sensors are used to display working conditions in all aspects for the real subway train bogies. The orinal datasets contain signals collected from 21 channels and the sampling frequency is 64kHZ. Considering that the original signals are relationships bewteen the amplitude and time, which is hard to recogonize what sort of discipline it obeys to. 
Meanwhile, we've found that the periodicity of the given datasets and the amount of the samples are limited. Thus, we split the data into ten parts with respect to the temporal scale for entire training process.
Thus, it is crutial to introduce several techniques to extract the characteristics of the data.

The wavelet transform is a signal processing technique that decomposes a signal into different frequency bands through multi-resolution analysis to extract feature information[\cite {mallat1989theory}]. The Morlet wavelet transform[\cite {lin2000feature}], a commonly used wavelet transform, combines a Gaussian window function with a sinusoidal wave, making it suitable for time-frequency analysis of non-stationary signals. 
The wavelet analysis is essentially represents signals using a series of finite basis functions, highlighting the instantaneous characteristics of signals that Fourier analysis, which uses trigonometric function bases, cannot express. 
The wavelet coefficient of a signal, $W_x(a,t)$, can be obtained through the convolution of the mother wavelet function $varphi(t)$ with the signal $x(t){:}$

\begin{equation}
W_x(a,b)=\frac{1}{\sqrt{a}}\int_{-\infty}^{+\infty}x(b)\varphi^*(\frac{t-b}{a})db
\label{euq:9}
\end{equation}
where $a$ is the scale of the wavelets and $b$ represents the local temporal orgin.

The commonly used Morlet wavelet takes the form of:[\cite{goupillaud1984cycle}]

\begin{equation}
\psi(t)=\pi^{-\frac14}e^{j\omega_0t}e^{-\frac{t^2}2}
\label{euq:10}
\end{equation}

where $\psi(t)$ is the Morlet wavelet function and $omega_0$ is the central frequency.

As illustrated in [\cite {liang2022intelligent}], the morlet wavelet basis function performs best among the 5 distinct wavelet basis functions, including Haar, Db4, Morlet, Coif5 and Sym8. Consequently, the morlet wavelet is adopted herein for feature extraction.

Noteworthy is the fact that the clear the image is, more information is therefore contained. However, to achieve better performance, the efficiency may decline. Thus, it's imperative to make a trade-off. As shown in [\cite{liang2022intelligent}], the 64*64 images resolutions may be a good choice for its efficiency and considerable accuracy, which is adopted for our settings.

After the wavelet transform, the original time-amplitude relation is tranformed to the feature images. Then, the datasets of the feature signals provided for training as well as for testing are segmented into the training set, the validation set which accounts 20 percentage of the total training set and the test set. 
\begin{figure}[h]
	\centering
	\includegraphics[width=0.8\textwidth]{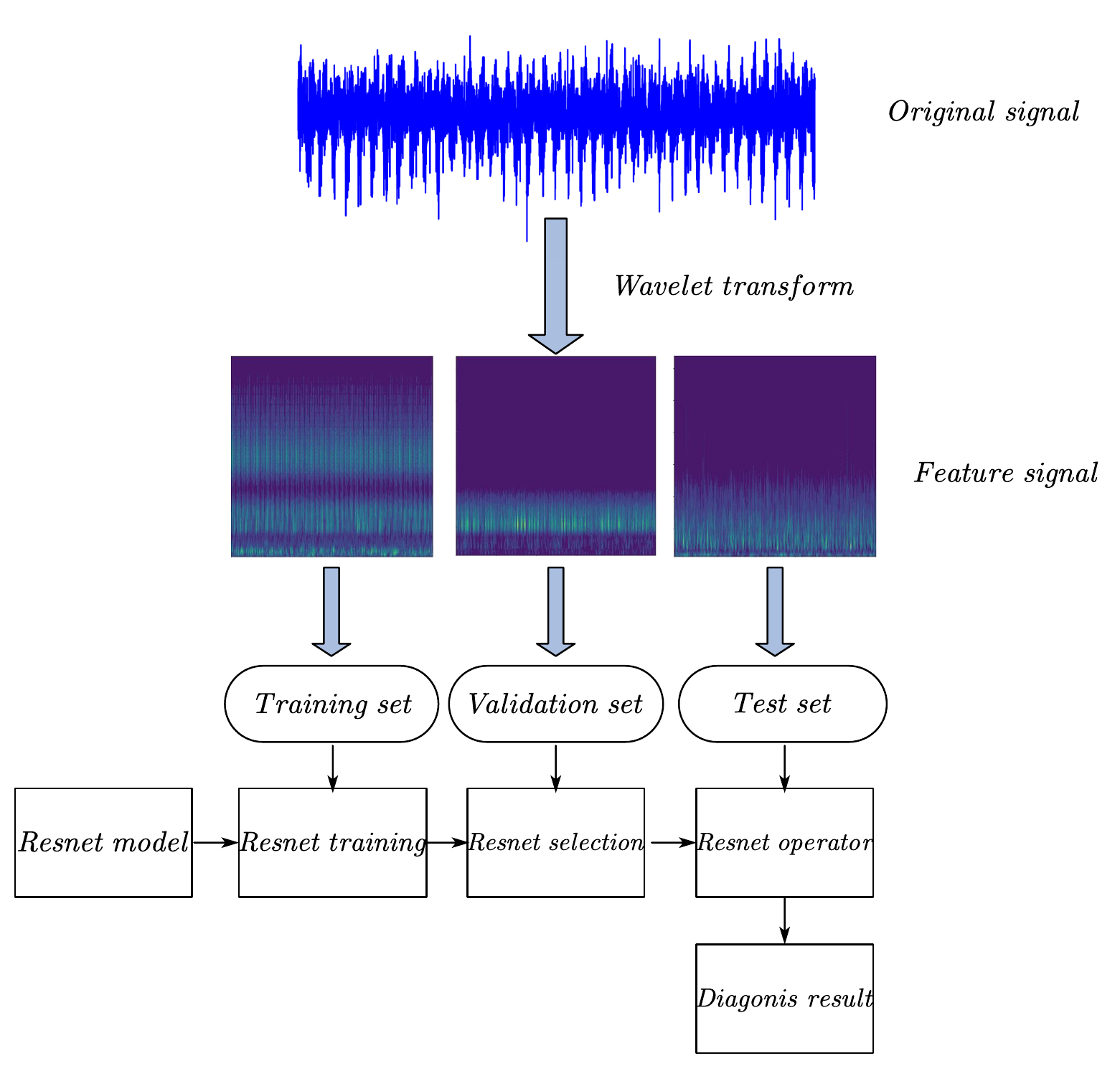}
	\caption{The overall framework of this work}
	\label{fig:illustrain}
\end{figure}
\section{Experiment}
\subsection{Case 1: Data analysis using the ResNet50 model}
\begin{figure}[h]
	\centering
	\includegraphics[width=0.8\textwidth]{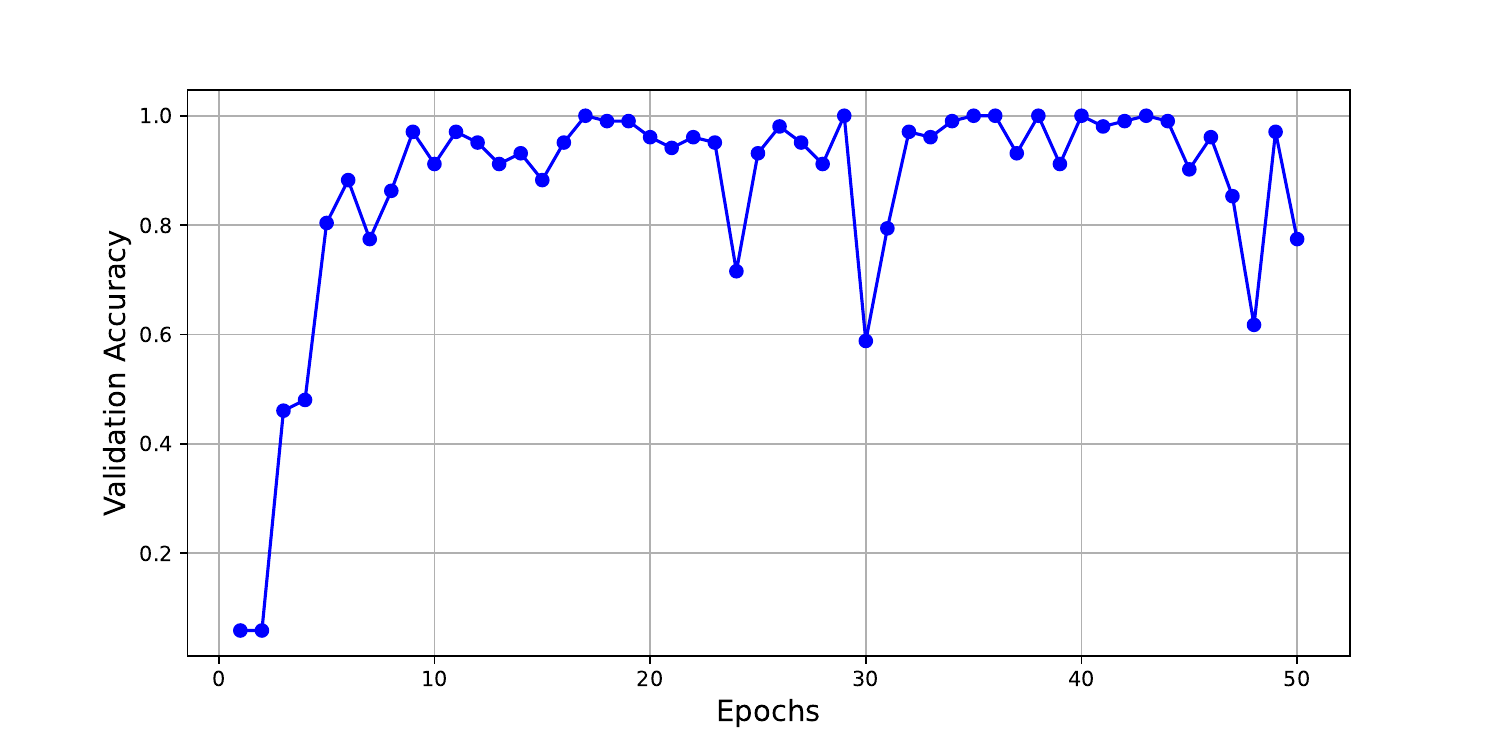}
	\caption{Resnet 50 Validation Accuracy over Epochs}
	\label{fig:3-1}
\end{figure}

\begin{figure}[h]
	\centering
	\includegraphics[width=0.8\textwidth]{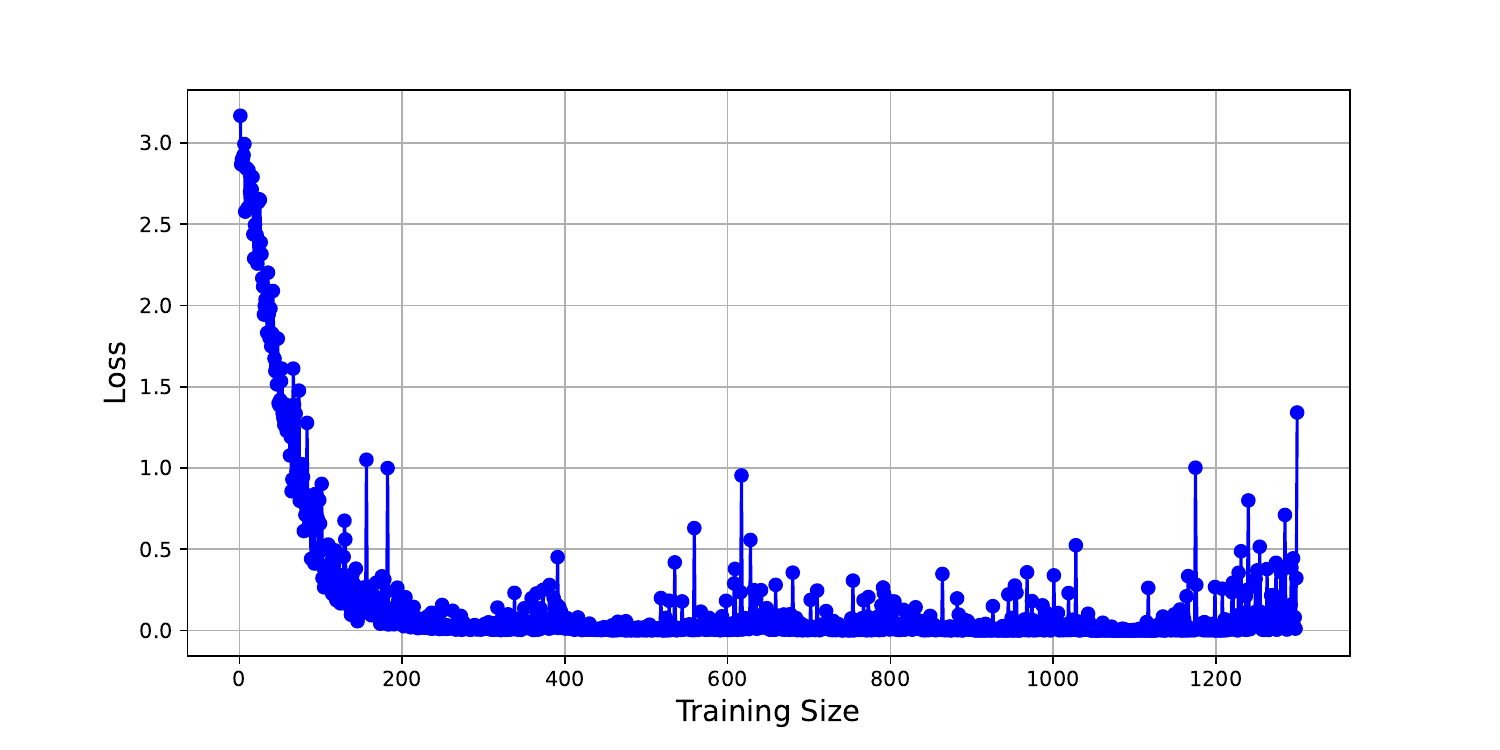}
	\caption{Resnet 50 Loss over Training Size}
	\label{fig:3-2}
\end{figure}

The validation accuracy over epochs, illustrated in Figure \ref{fig:3-1}, shows a rapid increase in the initial stages, quickly reaching approximately 0.8, indicating effective learning early on. This is followed by fluctuations between 0.8 and 1.0, maintaining a generally high level, suggesting good generalization. However, several significant drops, particularly around the $20th$ and $30th$ epochs, indicate potential issues with the model's ability to generalize in certain scenarios. Despite these fluctuations, the validation accuracy remains above 0.9 for most epochs, indicating relatively stable performance. To mitigate these fluctuations and improve generalization, one might consider using a larger validation set or implementing early stopping techniques to prevent overfitting.
The training loss over training size, depicted in Figure \ref{fig:3-2}, shows a rapid decrease from approximately 3.0 to below 0.1 in the initial stages, indicating effective error reduction and improved training accuracy. As training progresses, the loss stabilizes at a low level, suggesting gradual convergence. However, several spikes in the loss value, particularly around training sizes of 600 and 1200, suggest potential anomalies in specific data batches. These spikes indicate possible overfitting or underfitting, which could be mitigated by increasing dataset diversity or applying data augmentation techniques. Additionally, adjusting the learning rate or experimenting with different optimization algorithms might further smooth the loss curve. Overall, the model demonstrates strong performance, with high validation accuracy and low loss values, indicating effective learning and good generalization. Further optimization can be achieved by adjusting training parameters, enhancing dataset diversity, or employing advanced regularization techniques to improve stability and generalization.

\begin{figure}[h]
	\centering
	\includegraphics[width=0.8\textwidth]{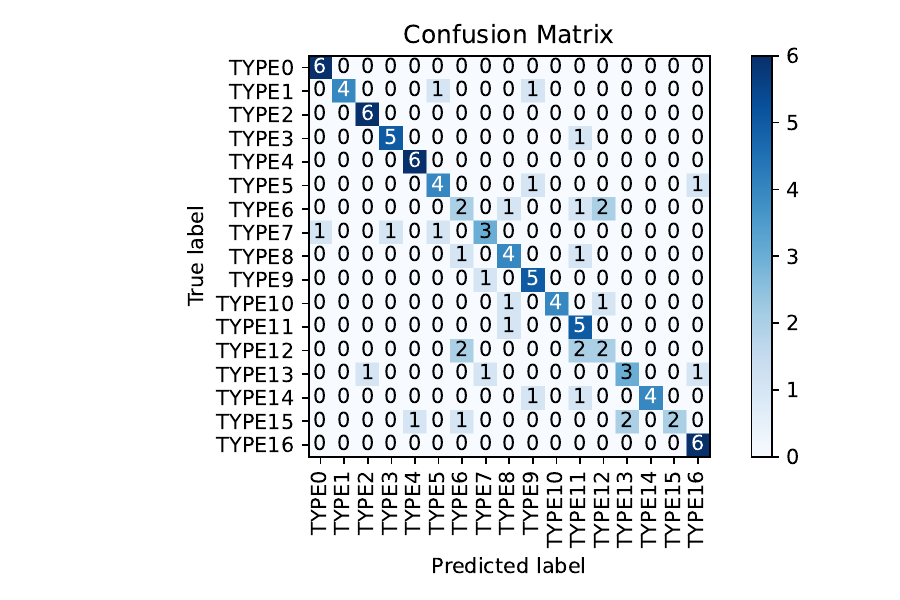}
	\caption{Confusion matrix of ResNet 50 }
	\label{fig:3-3}
\end{figure}

The confusion matrix, presented in Figure \ref{fig:3-3}, provides a detailed overview of the model's performance across different classes. The accuracy of the confusion matrix is 69.61$\%$. Table \ref{table:1} below showcases the precision, recall, and F1 Score for each class. The metrics are calculated according to the following formulas. The diagonal elements of the matrix are relatively high, indicating that the model correctly classifies a significant number of instances for most classes. For example, classes TYPE0, TYPE16, TYPE2, and TYPE4 have perfect or near-perfect classification with no misclassifications. However, there are some notable off-diagonal elements indicating misclassifications. For instance, TYPE6 has instances misclassified mainly as TYPE12. TYPE12 has instances misclassified mainly as TYPE6, suggesting some confusion between these classes and indicating potential overlaps in the features of these classes. Despite the presence of some misclassifications, the confusion matrix overall indicates that the model performs well across most classes. The majority of predictions fall along the diagonal, showing that the model has learned to distinguish between different classes effectively. To further improve the model's performance, future work could involve detailed analysis of the misclassified instances to understand the underlying reasons for confusion, and exploring advanced techniques such as ensemble methods to improve predictive performance. In summary, the overall evaluation suggests that while the model performs well, targeted enhancements could further boost its accuracy and robustness. 

\begin{equation}
\mathrm{Accuracy}=\frac{\mathrm{TP}+\mathrm{TN}}{\mathrm{TP}+\mathrm{TN}+\mathrm{FP}+\mathrm{FN}}
\label{euq:3-1}
\end{equation}

\begin{equation}
\mathrm{Precision}_i=\frac{\mathrm{TP}_i}{\mathrm{TP}_i+\mathrm{FP}_i}
\label{euq:3-2}
\end{equation}

\begin{equation}
\mathrm{Recall}_i=\frac{\mathrm{TP}_i}{\mathrm{TP}_i+\mathrm{FN}_i}
\label{euq:3-3}
\end{equation}

\begin{equation}
\mathrm{F1~Score}_i=2\times\frac{\mathrm{Precision}_i\times\mathrm{Recall}_i}{\mathrm{Precision}_i+\mathrm{Recall}_i}
\label{euq:3-4}
\end{equation}

\begin{table}[h]
\centering
\begin{tabular}{|c|c|c|c|}
\hline   Class    &   Precision  &  Recall  & F1 Score  \\
\hline   TYPE0  & 0.8571 & 1.0000 & 0.9231 \\
\hline   TYPE1  & 1.0000 & 0.6667 & 0.8000 \\
\hline   TYPE2  & 0.8571 & 1.0000 & 0.9231 \\
\hline   TYPE3  & 0.8333 & 0.8333 & 0.8333 \\
\hline   TYPE4  & 0.8571 & 1.0000 & 0.9231 \\
\hline   TYPE5  & 0.6667 & 0.6667 & 0.6667 \\
\hline   TYPE6  & 0.3333 & 0.3333 & 0.3333 \\
\hline   TYPE7  & 0.6000 & 0.5000 & 0.5455 \\
\hline   TYPE8  & 0.5714 & 0.6667 & 0.6154 \\
\hline   TYPE9  & 0.6250 & 0.8333 & 0.7143 \\
\hline   TYPE10 & 1.0000 & 0.8333 & 0.8000 \\
\hline   TYPE11 & 0.4545 & 0.3333 & 0.5882 \\
\hline   TYPE12 & 0.4000 & 0.5000 & 0.5436 \\
\hline   TYPE13 & 0.6000 & 0.6667 & 0.8000 \\
\hline   TYPE14 & 1.0000 & 0.3333 & 0.5000 \\
\hline   TYPE15 & 1.0000 & 1.0000 & 0.8571 \\
\hline   TYPE16 & 0.7500 & 1.0000 & 0.8571 \\
\hline
\end{tabular}
\caption{ResNet 50 Classification Evaluation Metrics}
\label{table:1}
\end{table}

%%%第二部分哩
\subsection{Case 2: Data analysis using the ResNet34 model}

\begin{figure}[h]
	\centering
	\includegraphics[width=0.8\textwidth]{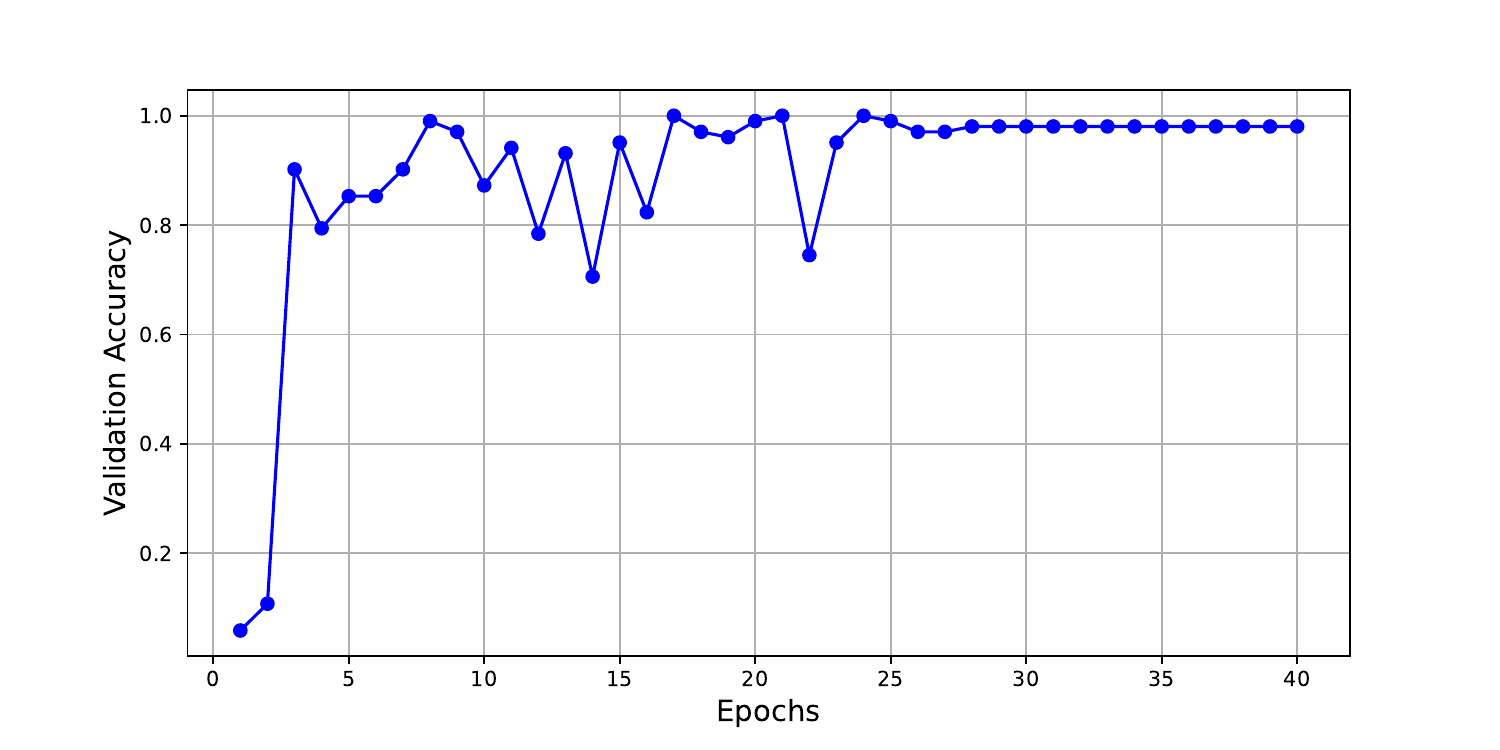}
	\caption{Resnet 34 Validation Accuracy over Epochs}
	\label{fig:3-4}
\end{figure}

The validation accuracy over epochs for ResNet34, shown in Figure \ref{fig:3-4}, demonstrates significant fluctuations between 0.7 and 1.0 for the first 25 epochs, indicating variability in performance during early training stages. However, after the 25th epoch, the validation accuracy stabilizes, showing minimal fluctuations and hovering around 0.98, indicating a high level of accuracy and stability in the later stages of training. This suggests that the validation accuracy likely converges after 25 epochs. This rapid stabilization and high accuracy suggest that ResNet34 is capable of learning effectively and generalizing well after initial variability.

\begin{figure}[h]
	\centering
	\includegraphics[width=0.8\textwidth]{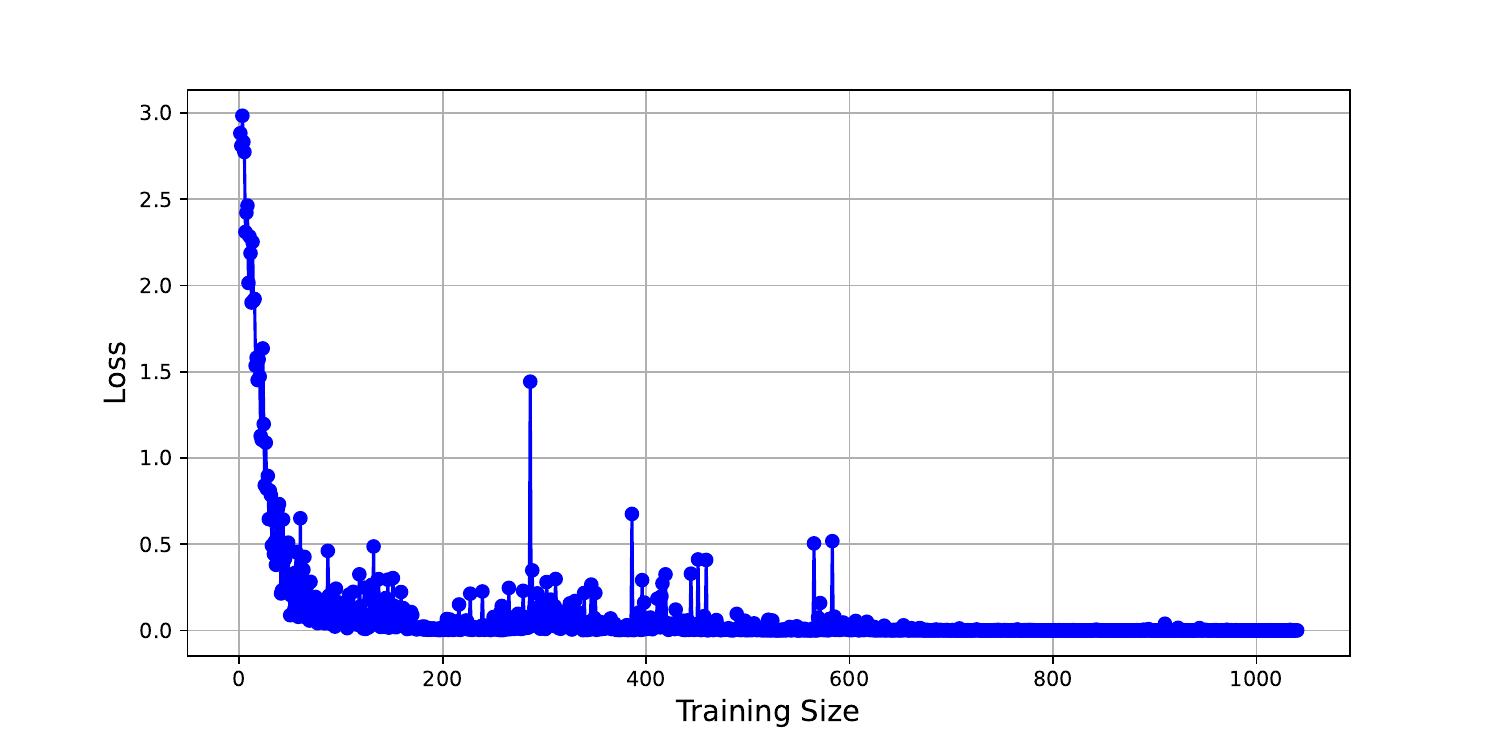}
	\caption{Resnet 34 Loss over Training Size}
	\label{fig:3-5}
\end{figure}

The training loss for ResNet34, depicted in Figure \ref{fig:3-5}, decreases sharply within the first 200 training samples, dropping below 0.01, which is faster than ResNet50. Beyond 600 training samples, the loss remains almost consistently below 0.1, indicating effective learning and low error rates. This suggests that the optimal number of samples might be around 600. However, occasional spikes in the loss value at training sizes around 280, 380, and 580 indicate potential anomalies or outliers in the training data. These spikes suggest that while the model learns quickly and efficiently, there are specific instances where it struggles, possibly due to irregularities in the data.

\begin{figure}[h]
	\centering
	\includegraphics[width=0.8\textwidth]{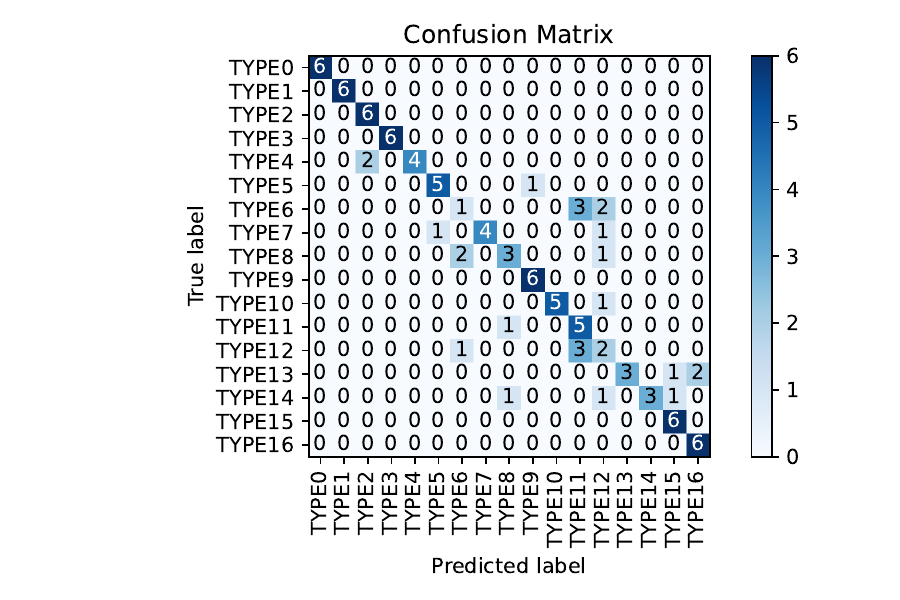}
	\caption{Confusion matrix of ResNet 34}
	\label{fig:3-6}
\end{figure}

The confusion matrix for ResNet34, shown in Figure \ref{fig:3-6}, indicates strong classification performance with most values concentrated along the diagonal. The accuracy of the confusion matrix is 75.49$\%$. Table\ref{table:2}  below showcases the precision, recall, and F1 Score for each class. The metrics are calculated according to the consistent formulas above. Classes such as TYPE0, TYPE1, TYPE2, TYPE3, TYPE9, TYPE15, and TYPE16 have perfect classification scores, reaching a value of 6, while TYPE5, TYPE10 and TYPE11 reach 5. This indicates that ResNet34 has a high accuracy for most classes. However, some classes exhibit notable misclassifications. For example, TYPE6 is often misclassified as TYPE11 and TYPE12, with only one correct classification, and TYPE12 frequently misclassified as TYPE11, indicating difficulty in distinguishing between these classes. Despite these misclassifications, ResNet34 shows overall improved accuracy and robustness compared to ResNet50, as evidenced by the higher concentration of correct classifications and fewer widespread misclassifications. In summary, while ResNet34 performs exceptionally well for most classes, further analysis and targeted adjustments are needed to address specific classification challenges, particularly among closely related classes.

\begin{table}[h]
\centering
\begin{tabular}{|c|c|c|c|}
\hline   Class  &   Precision  &  Recall  & F1 Score  \\
\hline   TYPE0  & 1.0000 & 1.0000 & 1.0000 \\
\hline   TYPE1  & 1.0000 & 1.0000 & 1.0000 \\
\hline   TYPE2  & 0.7500 & 1.0000 & 0.8571 \\
\hline   TYPE3  & 1.0000 & 1.0000 & 1.0000 \\
\hline   TYPE4  & 1.0000 & 0.6667 & 0.8000 \\
\hline   TYPE5  & 0.8333 & 0.8333 & 0.8333 \\
\hline   TYPE6  & 0.2500 & 0.1667 & 0.2000 \\
\hline   TYPE7  & 1.0000 & 0.6667 & 0.8000\\
\hline   TYPE8  & 0.6000 & 0.5000 & 0.5455\\
\hline   TYPE9  & 0.8571 & 1.0000 & 0.9235 \\
\hline   TYPE10 & 1.0000 & 0.8333 & 0.9091 \\
\hline   TYPE11 & 0.4545 & 0.8333 & 0.5882 \\
\hline   TYPE12 & 0.2500 & 0.3333 & 0.2857 \\
\hline   TYPE13 & 1.0000 & 0.5000 & 0.6667 \\
\hline   TYPE14 & 1.0000 & 0.5000 & 0.6667 \\
\hline   TYPE15 & 0.7500 & 1.0000 & 0.8571 \\
\hline   TYPE16 & 0.7500 & 1.0000 & 0.8571 \\
\hline
\end{tabular}
\caption{ResNet 34 Classification Evaluation Metrics}
\label{table:2}
\end{table}
\section{Conclusion}
The research concludes that the integration of wavelet transform and ResNet in a fault diagnosis model for urban rail transit systems yields promising results. The WT-ResNet model not only captures the transient nature of fault signals effectively but also learns complex representations of these signals for accurate classification. Future work should focus on extending the model to handle a wider range of fault types and improving its real-time application capabilities. The findings contribute to the advancement of intelligent diagnostics in the transportation sector, with potential applications beyond urban rail systems.

\section*{Declaration of Competing Interest }
The authors declare that they have no known competing financial interests or personal relationships that could have appeared to influence the work reported in this paper.

\newpage

\bibliographystyle{unsrtnat}
\bibliography{references}  %%% Uncomment this line and comment out the ``thebibliography'' section below to use the external .bib file (using bibtex) .

\end{document}